\documentclass[a4paper]{jpconf}
\usepackage{iopams}  
\usepackage{graphicx}
\usepackage{amssymb}
\usepackage{citesort}
\usepackage{caption}
\usepackage{amsfonts}
\usepackage{amssymb}
\usepackage{amsopn}
\usepackage{epsfig}
\usepackage{cite}
\usepackage{graphics,psfrag,rotating}
\usepackage{dcolumn}
\usepackage{bm}
\usepackage{epstopdf}
\usepackage{color}
\usepackage[usenames,dvipsnames,svgnames]{xcolor}
\usepackage[colorlinks=true,
            linkcolor=red,
            urlcolor=gray,
            citecolor=blue]{hyperref}
  \usepackage{hyperref}
   \usepackage{subfloat}
\usepackage{subfig}

\def\3nab{\tilde{\nabla}}

\def\p{\partial}

\def\hsp5{\hspace{5mm}}
\newcommand{\sfrac}[2]{{\textstyle{#1\over#2}}}
\def\case#1/#2{\textstyle\frac{#1}{#2}}

\def\ber {\begin{eqnarray}}
\def\eer {\end{eqnarray}}
\def\bea {\begin{eqnarray}}
\def\eea {\end{eqnarray}}

\def\bc {\begin{center}}
\def\ec {\end{center}}
\def\case#1/#2{\frac{#1}{#2}}

\newcommand{\bw}{\begin{widetext}}
\newcommand{\ew}{\end{widetext}}
\newcommand{\nn}{\nonumber\\}

\newcommand{\be}{\begin{equation}}
\newcommand{\bse}{\begin{subequation}}
\newcommand{\ese}{\end{subequation}}
\newcommand{\ee}{\end{equation}}
\newcommand{\eei}{\end{eqnarray}\indent\indent}
\newcommand{\ba}{\begin{array}}
\newcommand{\ea}{\end{array}}
\newcommand{\bal}{\begin{eqnarray}}
\newcommand{\eal}{\end{eqnarray}}

\def\case#1/#2{\textstyle\frac{#1}{#2} }
\newcommand{\nb}{\nabla}

\newcommand{\bver}{\begin{verbatim}}
\newcommand{\ever}{\end{verbatim}}

\newcommand{\hb}{\hbar}

\begin{document}
\nocite{*}

\title{Oscillating cosmological correlations in $f(R)$ gravity}


\author{Neo Namane, Heba Sami and Amare Abebe}

\address{Center for Space Research \& Department of Physics, North-West University, Mafikeng, South Africa}
\ead{nechnotick@gmail.com}

\begin{abstract}
The purpose of this paper is to investigate the oscillatory behavior of the universe through a Schr\"{o}dinger-like Friedmann equation and a modified gravitational background described by the theory of $f(R)$ gravity. The motivation for this stems from the observed periodic behaviour of large-scale cosmological structures when described within the scope of the general theory of relativity. The analysis of the modified Friedmann equation for the dust epoch in power-law $f(R)$ models results in different behaviors for the wave-function of the universe. 
\end{abstract}

\section{Introduction}

According to the Cosmological Principle (CP), when viewed on sufficiently large scales ($\sim 1000h^{-1}$ Mpc), matter in the universe is homogeneously and isotropically distributed. However, observations show that the CP breaks on scales of the order $\sim 100h^{-1}$ Mpc and below, and the clustering property of cosmological objects (galaxies, clusters, superclusters, filaments) shows there exists some sort of hierarchy.  Thus, there are suggestions that the distribution of galaxies is not random and that some fundamental mechanism has led to the formation of large-scale structure. One proposal is cosmological solutions with an overall Friedmannian expanding behaviour, corrected by small oscillatory regimes \cite{cap2000}.
Given the cosmological scale factor $a(t)$ and redshift $z$, we have the Hubble parameter $H$ given by
\be
\frac{\dot{a}}{a}=\frac{\Theta}{3}\equiv H=-\frac{\dot{z}}{1+z}\;.
\ee
Oscillations at a particular redshift can be considered as some sort of quantization \cite{cap2000,tif77} and all quantities containing $H$ or $z$ have to oscillate. These oscillations affect several observational quantities, such as the number count of galaxies
\be
\frac{dN}{d\Omega dLdz}=n(L,t_0)a^2_0H^{-1}d^2\;,
\ee
where $dN$ is the number of galaxies in the solid angle $d\Omega$ having redshift between $z$ and $z+dz$ and luminosity between $L$ and $L+dL$,$n(L,t_0)$ that represents the number density of galaxies with luminosity $L$ that an observer sees at time $t_0$,  $a_0$ is the value of the cosmological scale factor today and  $d$ is the comoving distance defined as $d=\int^{t_0}_t\frac{dt}{a}$.
\newline
There have also been recent attempts to link gravitation with quantization, largely motivated by the need to unify two of theoretical physics' most fundamental theories into one overarching framework.
There are generally two main approaches in this endeavor:
\begin{itemize}
\item The whole universe as a quantum system of co-existing and non-interacting universes \cite{wit67,ev57}
\item The universe as a classical background: where primordial quantum processes gave rise to the current macroscopic structures \cite{bir84}
\end{itemize}
Following Capozziello \cite{cap2000} and Rosen \cite{ros93}, one can recast the cosmological Friedmann equation
\be
\left(\frac{\dot{a}}{a}\right)^2=\frac{1}{9}\Theta^2=\frac{\mu}{3}-\frac{k}{a^2}
\ee
 as some sort of a  Schr\"{o}dinger equation (SE). To do so, we can  rewrite the above equation as the equation of motion of a ``particle" of mass m:
 \be\label{eomm}
 \frac{1}{2}m\dot{a}^2-\frac{m}{6}\mu a^2=-\frac{1}{2}mk\;,
 \ee
 where $\Theta\equiv 3H$, $\mu$ and $k$ are, respectively, the cosmological (volume) expansion parameter, the energy density and spatial curvature of the universe. The total energy $E$ of the particle can be thought of as being the sum of the kinetic $T$ and potential $V$ energies:
 \be
 E=T+V
 \ee
 where
\be
T=\frac{1}{2}m\dot{a}^2\;,\;\;\;\; V=-\frac{1}{6}m\mu a^2\;,\;\;\;\; E=-\frac{1}{2}mk\;.
\ee
One can also rewrite the Raychaudhuri (acceleration) equation
 \be
 \frac{\ddot{a}}{a}=-\frac{1}{2}(\mu+3p)
 \ee
 in a way that mimics the equation of motion of the particle, otherwise given by
 \be
 m\ddot{a}=-\frac{dV}{da}\;,
 \ee
 where $p$ is the isotropic pressure, related to the energy density of a perfect fluid through the equation of state parameter $w$ as
 $p=w\mu$.
 The particle's momentum and Hamiltonian are defined, respectively, as
 \be
 \Pi\equiv m\dot{a}\;,\;\;\;\;\ H\equiv \frac{\Pi^2}{2}+V(a)\;.
 \ee
 From the ``first quantization" scheme, we have
 \be
 \Pi\to -i\hbar\frac{\p}{\p a}\;.
 \ee
Thus the SE for the wavefunction $\Psi=\Psi(a,t)$ is given by
 \be
 i\hbar\frac{\p\Psi}{\p t}=-\frac{\hbar^2}{2m}\frac{\p^2\Psi}{\p a^2}+V(a)\Psi\;.
 \ee
We can think of $m$ as the mass of a galaxy, and $|\Psi|^2$  as the probability of finding the galaxy at $a(t)$ or at a given redshift 
 \be
 1+z=\frac{a_0}{a}\;,
 \ee
and thus, in the language of quantum physics,  $\Psi=\Psi(z,t)$ defines the probability amplitude to find a given object of mass $m$ at a given redshift $z$, at time $t$. The stationary states of energy $E$ are given by
 \be
 \Psi(a,t)=\psi(a)e^{-iEt/\hbar}\;,
 \ee
 and the time-independent Schr\"{o}dinger equation (TISE) reads
 \be\label{tise}
 -\frac{\hb^2}{2m}\frac{d^2\psi}{da^2}+V\psi=E\psi\;.
 \ee

\section{$f(R)$ Gravitation}
 $f(R)$ models are a sub-class of {\it fourth-order} theories of gravitation, with an action given by \cite{abebe2014anti} \footnote{In geometrized units: $c= 8\pi G\equiv1$.}
\be
{\cal A}_{f(R)}= \sfrac12 \int d^4x\sqrt{-g}\left[f(R)+2{\cal L}_m\right]\;,
\label{action}
\ee
where $R$, $g$ and ${\cal L}_{m}$  are the Ricci scalar, the determinant of the  metric tensor, and the matter Lagrangian. The $f(R)$-generalized Einstein field equations can be given by
 \be
 f'G_{ab}=T^{m}_{ab}+\sfrac{1}{2}(f-Rf')g_{ab}+\nb_{b}\nb_{a}f'-g_{ab}\nb_{c}
\nb^{c}f'\;. 
\ee
Here primes symbolize derivatives with respect to $R$, whereas $G_{ab}$ and $T^{m}_{ab}$  are the Einstein tensor and the energy-momentum tensor of matter respectively. 
These models provide the simplest generalizations to GR, and come with an extra degree of freedom. The cosmological viability of the models can be determined through observational and theoretical constraints. Some generic viability conditions on $f$ include \cite{amare2015beyond}:
\begin{itemize}
\item To ensure gravity remains attractive
\be f' > 0 ~~\forall R\ee
\item For  stable matter-dominated   and  high-curvature  cosmological regimes (nontachyonic scalaron) 
\be
f''>0~~\forall R\gg f''\ee
\item  GR-like law of gravitation in the early universe (BBN, CMB constraints)
\be
 \lim_{R \to \infty} \frac{f(R)}{R} = 1\Rightarrow f'<1
\ee
\item  At recent epochs
\be
| f'-1 | \ll 1
\ee
\end{itemize}

The matter-energy content of a universe filled with a perfect fluid is specified by 

\be T_{ab}=(\mu+p)u_{a}u_{b}+pg_{ab}\;.\ee

The background curvature and total perfect fluid thermodynamics is described by \cite{ntahompagaze2017f}
\ber
&&\label{mur}\mu_{R}=\frac{1}{f'}\left[\frac{1}{2}(Rf'-f)-\Theta f'' \dot{R}\;  \right],\nn
&&\label{pr}p_{R}=\frac{1}{f'}\left[\frac{1}{2}(f-Rf')+f''\ddot{R}+f'''\dot{R}^{2}+\frac{2}{3}\Theta f''\dot{R}\right]\;,\nn
&&\mu\equiv\frac{\mu_{m}}{f'}+\mu_{R}\;,~~p\equiv\frac{p_{m}}{f'}+p_{R}\;.
\eer

\section{The Cosmological Schr\"{o}dinger Equation}
In $f(R)$ gravity, the Raychaudhuri equation generalizes to
\begin{eqnarray}\label{Fri1}
&& \frac{\ddot{a}}{a}= -\frac{1}{6}\Big(\frac{\mu_{m}}{f^{'}}+ \mu_{R}+ \frac{3p_{m}}{f^{'}}+3p_{R}\Big)
\end{eqnarray}
which in terms of the expressions for $\mu_R$, $p_R$ and the trace equation ($R=\mu-3p$) can be simplified as
\begin{equation}\label{eq:14}
\frac{\ddot{a}}{a}= -\frac{1}{6f^{'}}\Big( 2\mu_{m}-f -2\Theta f^{''}\dot{R}\Big)\;.
\end{equation}
Similarly, the corresponding  modified Friedmann equation in $f(R)$ gravity is given by
\begin{equation}\label{Ray1}
\frac{1}{9}\Theta^2+\frac{k}{a^{2}}= \frac{1}{6f^{'}}\Big(2\mu_{m}+ Rf^{'}-f-2\Theta f^{''}\dot{R}\Big)\;.
\end{equation}
But for FLRW models, it is also true that
\begin{equation}\label{Ray11}
\frac{1}{9}\Theta^2+\frac{k}{a^{2}}= \frac{1}{6}R- \frac{\ddot{a}}{a}\;, 
\end{equation}
one can therefore re-write equation \eref{eomm} as
\begin{equation}\label{Ray2}
 \frac{1}{2}m\dot{a}^{2}-\frac{1}{2}\Big( \frac{R}{6}-\frac{\ddot{a}}{a}\Big)ma^{2}=-\frac{1}{2}mk\;,
\end{equation}
with the potential
\begin{equation}\label{vv}
 V(a) = -\frac{1}{2} \Big( \frac{R}{6}-\frac{\ddot{a}}{a}\Big)ma^{2}\;.
\end{equation}
Now rearranging the TISE (\ref{tise}) for $f(R)$ gravity yields 
\be\label{psieq}
 \frac{d^{2}\psi}{da^{2}}= -\frac{2m}{\hbar^{2}}\Big[E-V(a)\Big]\psi= \Big[C-\frac{m^{2}}{6f^{'}\hbar^{2}}\Big( 2\mu_{m}+Rf^{'}-f-2\Theta f^{''}\dot{R}\Big)a^{2}\Big]\psi\;,
\ee
where $C\equiv \frac{m^{2}k}{\hbar^{2}}$. 
\section{Oscillating Solutions}

Let us now consider power-law $f(R)$ models of the form
\begin{equation}
f(R)= R^{n}\;,
\end{equation} 
admitting  scale factor solutions 
\begin{equation}
a(t)= a_0t^{\frac{2n}{3(1+w)}}\;.
\end{equation}
 For dust ($w=0)$ models, we get
\be\label{expansion}
\Theta= \frac{2n}{t}\;,~~~
R=\frac{4n(4n-3)}{3t^{2}}\;,~~
\mu_{m}= \frac{\mu_0}{a^{3}}\;.
\ee
Here $a_0$ and $\mu_0$ are integration constants that can be normalized to unity when considering current values of the scale factor and the energy density of matter.
Thus, for such models, the TISE (\ref{psieq}) takes the form
\ber\label{psii}
&&\frac{d^{2}\psi}{da^{2}} = \left\{ C- \frac{m^{2}}{3n\hbar^{2}a^{3}}\left[\frac{4n(4n-3)}{3a^{\frac{3}{n}}}\right]^{1-n} + \frac{4nm^{2}(4n-3)}{18\hbar^{2}a^{\frac{3-2n}{n}}}\right. \nn
&&\left.~~~~~~- \frac{4m^{2}(4n-3)}{18\hbar^{2}a^{\frac{3-2n}{n}}} + \frac{32nm^{2}(n-1)}{24a^{\frac{3-2n}{n}}}\right\}\psi\;.
\eer
For $n=1$, the above equation reduces to
\begin{equation}
\frac{d^{2}\psi}{da^{2}}= \Big[C-\frac{B}{a}\Big]\psi\;,
\end{equation}
where $B\equiv\frac{m^{2}}{3\hbar^{2}}$ and we recover the GR solutions\cite{cap2000} obtained by Capozziello et al. For example, for a flat universe, $C=0$ and we get a combination of Bessel functions as the general solution:
\be
\psi \left( a \right) ={\it C1}\,\sqrt {a}{{\sl J}_{1}\left(2\,
\sqrt {-B}\sqrt {a}\right)}+{\it C2}\,\sqrt {a}{{\sl Y}_{1}\left(2\,
\sqrt {-B}\sqrt {a}\right)}\;.
\ee
For $n=3$, it can be shown that
\begin{equation}
 \frac{d^{2}\psi}{da^{2}}= \Big[C - Ba\Big]\psi\;,
\end{equation}
and the corresponding solutions are Airy functions of the form
\be
\psi \left( a \right) ={\it C3}\,{{\rm Ai}\left({\frac {C-Ba}{{B}^{
2/3}}}\right)}+{\it C4}\,{{\rm Bi}\left({\frac {C-Ba}{{B}^{2/3}}}
\right)}\;.
\ee
Figure \eref{fig1} below shows the oscillatory behaviour of such exact-solution wavefunctions. Our investigations for the power-law models suggest that exact solutions are possible only for $n=1$ and $n=3$.  Also, our numerical computations show no oscillatory behaviour of the solutions for $n<1$, whereas oscillating solutions for $n=1.5$ and $n=2$ are presented in figure \eref{fig2}.
\begin{figure}[h!]
  \centering
  \subfloat[Oscillating wavefunction for $n=1$.]{\includegraphics[width=0.5\textwidth]{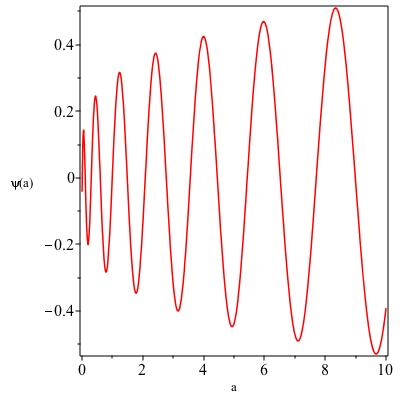}\label{fig1}}
  \subfloat[Oscillating wavefunction for $n=3$.]{\includegraphics[width=0.5\textwidth]{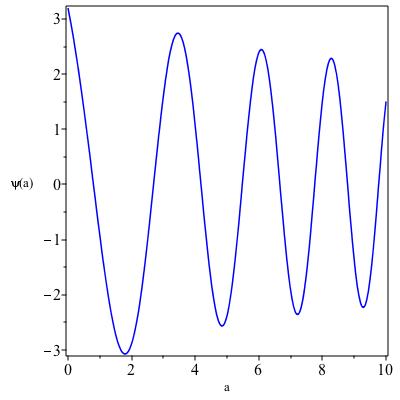}\label{fig3}}
  \caption{Exact solutions for $n=1$ and $n=3$.}
      \label{fig1}
\end{figure}
\unskip
\begin{figure}[h!]
  \centering
  \subfloat[Oscillating wavefunction for $n=1.5$.]{\includegraphics[width=0.5\textwidth]{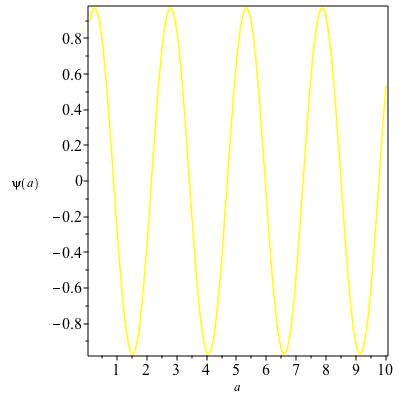}\label{fig1-5}}
  \subfloat[Oscillating wavefunction for $n=2$.]{\includegraphics[width=0.5\textwidth]{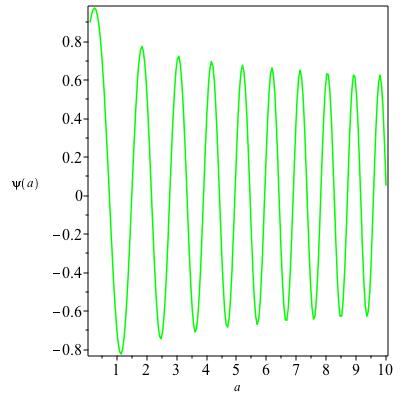}\label{2}}
    \caption{Numerical solutions for $n=1.5$ and $n=2$.}
      \label{fig2}
\end{figure}
\newpage
\section{Conclusion}
A breaking of homogeneity and isotropy on small scales with oscillating correlations between galaxies can be achieved with a Schr\"{o}dinger-like equation.
This work reproduces existing GR solutions and provides an even richer set of solutions for $f(R)$ gravity models, thus providing possible constraints on such models using observational data. For the power-law $f(R)$ model considered in this work, exact solutions have been obtained for $n=1$ and $n=3$ in the flat FLRW background, as well as numerical solutions for the $n=1.5$ and $n=2$ dust scenarios. A more detailed analysis of such oscillatory solutions with more viable $f(R)$ models and under more realistic initial conditions is currently underway.

\ack NN and HS acknowledge the Center for Space Research of North-West University for financial support to attend the  63rd Annual Conference of the South African Institute of Physics. NN acknowledges funding from the National Institute of Theoretical Physics (NITheP). HS and AA acknowledge that this work is based on the research supported in part by the 
National Research Foundation (NRF) of South Africa.

\section*{References}
\bibliography{ref}
\bibliographystyle{iopart-num}
\end{document}